\documentclass[iop]{emulateapj}
\usepackage{amsmath}
\usepackage{url}
\usepackage[T1]{fontenc}
\shorttitle{The Double Contact Nature of TT Herculis}
\shortauthors{Terrell \& Nelson}

\begin{document}

\title{The Double Contact Nature of TT Herculis}

\author{Dirk Terrell}
\affil{Department of Space Studies, Southwest Research Institute, 1050 Walnut St., Suite 300,
    Boulder, CO 80302}
\email{terrell@boulder.swri.edu}
\and
\author{Robert H. Nelson}
\affil{1393 Garvin Street, Prince George, BC, Canada, V2M 3Z1}
\email{bob.nelson@shaw.ca}

\begin{abstract}
We present new radial velocities and photometry of the short-period Algol TT Herculis. Previous attempts to model the light curves of the system have met with limited success, primarily because of the lack of a reliable mass ratio. Our spectroscopic observations are the first to result in radial velocities for the secondary star, and thus provide a spectroscopic mass ratio. Simultaneous analysis of the radial velocities and new photometry shows that the system is a double contact binary, with a rapidly rotating primary that fills its limiting lobe.
\end{abstract}


\section{Introduction}

TT Herculis (GSC 1521-00071, Hipparcos 82710) is a close binary star with an orbital period of 0.91 days. The Tycho catalog \citep{tycho} lists its $V_T$ magnitude as $9.72\pm 0.02$, making it a reasonably bright system, and the Hipparcos parallax, based on the improved reduction of the raw data by \cite{vanleeuwen} is $1.12\pm 1.04$ mas, placing it at a distance of $900 \pm 800$ pc. The variability of the system has been known for over a century \citep{lui10}. \cite{sel06} and \cite{mil89} discuss the early observational history of the system. The light curve shows a large difference in the eclipse depths, and between eclipses the light curve varies continuously, indicating large distortions of the shapes of the stars. Although the system has been the subject of numerous studies, there has been little agreement on even fundamental properties such as the spectral type, or the morphological type of the system, with previous light curve solutions running the gamut from detached to overcontact configurations.  \cite{san37} lists the spectral type of the system as A0. \cite{baldwin39} cites private communication with A. J. Cannon who classified it as A7. \cite{adams35} classify it as A3. \cite{hill75} give two values at different orbital phases, A7 and F2.

\cite{san37} published 18 radial velocities for the primary component and claimed to see doubling of lines on a few spectra, indicating that the mass ratio was slightly less than 0.5. However, he thought the evidence was too weak to publish the secondary's radial velocities. He classified the spectrum of the primary as A0 and cautiously noted that his observations near primary minimum showed the Rossiter effect \citep{ros24}. These are the only known radial velocities of the system until now.

Several light curves of the system have been published. \cite{hogg55} published photoelectric light curves in blue and yellow filters and, not surprisingly, were unable to find a satisfactory fit to the light curve using the Russell model. \cite{vangen62} published another photoelectric light curve taken with a blue filter. \cite{land68} published the first light curve data on the $UBV$ system but only covered the primary minimum. \cite{burchi82} published extensive $UBV$ observations. \cite{kwee83} published $BV$ observations made in 1962 and 1963, and were the first to analyze the light curves using a program capable of properly modeling the distorted shapes of the stars, the Wilson-Devinney program \citep[][hereafter, WD]{rew79,wd71}.

A reasonably good set of times of minimum for the system are available. Recent studies of the eclipse timing diagram find both periodic and secular terms in the changing orbital period. \cite{sel06} and \cite{kre08} both interpret the periodic terms as arising from a low mass third star orbiting the eclipsing pair with a period of approximately forty years, and the secular term as arising from mass transfer from the more massive to the less massive component. \cite{sel06} also discuss the possibility that the periodic terms are due to the Applegate mechanism \citep{app92} but cite the lack of observational data on brightness variations to be able to confirm that hypothesis.

\section{Observations}

In April of 2005, 2006 and 2007 R.H.N. took a total of 9 medium resolution (reciprocal dispersion = 10 \AA/mm, R = 10,000) spectra at the Dominion Astrophysical Observatory (DAO) in Victoria, BC using the 1.8m Plaskett telescope. The exposure time was usually one hour and the spectral coverage was approximately 5000-5260 \AA. Intermediate reductions (overscan removal, cosmic ray cleaning, setting apertures, fitting background, summation of counts, reduction to 1 dimension, calibration from Fe-Ar arc spectra, and finally dispersion correction) were performed by `Ravere', software developed by R.H.N. (Nelson, 2010). Final determination of radial velocities was performed by ``Broad'', software developed by the same author that uses the Rucinski broadening functions \citep{ruc02}. One spectrum was taken at phase 0.05 and did not result in a reliable radial velocity for the primary. The measured velocity was 13 km\: sec$^{-1}$ while the predicted value should be about -40 km\: sec$^{-1}$. We thus decided that the measured value was unreliable and excluded it from further consideration. The log of the spectroscopic observations is given in Table 1. Our light and velocity curve solutions all employed the time of the observations, rather than orbital phase, as the independent variable, and thus we solved for the reference epoch ($HJD_0$) and the orbital period ($P$). We take the values for those parameters for the solution with an early-type primary discussed in section \ref{sec:solution}, and all orbital phases given in Table 1 and elsewhere are computed with the values from that solution.

\begin{deluxetable*}{cccccc}
\tablecaption{Log of Spectroscopic Observations}
\tablenum{1}
\tablehead{
   \colhead{DAO} & \colhead{Mid Time} & \colhead{Orbital} & \colhead{Exposure} & \colhead{$V_1$} & \colhead{$V_2$}\\ 
   \colhead{Image Number} & \colhead{(HJD-2400000)} & \colhead{Phase} & \colhead{($sec$)} & \colhead{($km\: sec^{-1}$)} & \colhead{($km\: sec^{-1}$)}
} 
\startdata
3183 & 53489.8640 & 0.773 & 3600 & $114.3\pm5.7$ & $-197.1\pm6.3$ \\
3653 & 53842.9154 & 0.884 & 3600 & $114.9\pm5.9$ & $-165.6\pm2.7$ \\
3694 & 53847.9123 & 0.327 & 3600 & $-40.8\pm2.4$ & $243.0\pm2.6$ \\
3700 & 53847.9827 & 0.404 & 3600 & $-16.5\pm2.5$ & --- \\
3702 & 53848.0255 & 0.451 & 3600 & $0.1\pm2.0$ & --- \\
3769 & 53850.9415 & 0.648 & 3600 & $95.6\pm1.1$ & $-173.2\pm5.0$ \\
3771 & 53850.9839 & 0.695 & 3600 & $110.7\pm1.5$ & $-187.2\pm2.8$ \\
1355 & 54163.0471 & 0.832 & 1201 & $131.1\pm3.6$ & $-174.9\pm3.8$ \\
\enddata
\label{rvtable}
\end{deluxetable*}

Photometric observations in the $VR_CI_C$ photometric passbands were undertaken by R.H.N. at his private observatory in Prince George, BC, Canada in May and June of 2008.  A total of 392 frames in $V$, 413 in $R_C$ and 393 in $I_C$ were taken. Standard reduction techniques (bias subtraction, dark subtraction, and flatfielding) were done and aperture photometry performed. Differential photometry was then performed with GSC 1525-0805 ($V=9.82\pm0.03$ and $B-V=0.72\pm0.03$ from the Tycho-2 catalog \citep{tycho}) as the comparison star and GSC 1525-0939 ($V=11.73\pm0.07$ and $B-V=0.82\pm0.07$ from Data Release 7 of the APASS survey \citep{apass}) as the check star. No variation in the comparison and check stars greater than 0.01 magnitudes was detected, and the precision of the differential magnitudes for TT Her was about 0.005 magnitudes in each filter . The instrumental magnitude differences are given in Table 2.

\begin{deluxetable*}{ccccccccc}
\tablecaption{Photometry of TT Herculis}
\tablenum{2}
\tablehead{
   \colhead{HJD-2400000.0} & \colhead{Phase} & \colhead{$\Delta V$} & \colhead{HJD-2400000.0} & \colhead{Phase} & \colhead{$\Delta R_C$} & \colhead{HJD-2400000.0} & \colhead{Phase} & \colhead{$\Delta I_C$} 
} 
\startdata
54602.72516&0.882&-0.112&54602.72385&0.881& 0.174&54602.75011&0.910& 0.586\\
54602.72990&0.888&-0.099&54602.72587&0.883& 0.170&54602.75278&0.913& 0.597\\
54602.73256&0.891&-0.108&54602.72818&0.886& 0.193&54602.75544&0.916& 0.613\\
54602.73522&0.893&-0.095&54602.73061&0.888& 0.192&54602.75810&0.919& 0.629\\
54602.73788&0.896&-0.086&54602.73328&0.891& 0.210&54602.76076&0.921& 0.635\\
\enddata
\tablecomments{Table 2 is published in its entirety in the electronic edition of the Astrophysical Journal. A portion is shown here for guidance regarding its form and content.}
\label{ptmtable}
\end{deluxetable*}

\section{Analysis}
\subsection{Spectral Type of the Primary}

 Given the short orbital period and substantial proximity effects in TT Her, it is not too surprising that there has been disagreement over the spectral type of the system, but the values do seem unusually disparate. Clearly the classification of the TT Her spectrum is a difficult task, given that several experienced and highly regarded spectroscopists have arrived at such different values. An extended spectroscopic study of the system over several seasons to determine, for example, whether the spectral type is variable on longer timescales or even over the orbital period, might prove rewarding in solving the mystery of the wide variation of the spectral type determinations.

Another approach to estimating the temperature of the primary, which is needed to analyze the light curves, is to use the observed colors of the system. The amount of interstellar reddening then obviously plays a crucial role. \cite{land68} found $B-V=0.32$ and estimated $E(B-V)=0.1$, leading to an A8 classification. \cite{hilditch75} observed TT Her on the Str\"{o}mgren system, thus allowing us to determine the reddening in the $b-y$ color using the observed quantity $c_1=(u-v)-(v-b)$. \cite{shob84} gives a polynomial representation of the relationship between the intrinsic color $(b-y)_0$ and the intrinsic quantity $c_0=(u-v)_0-(v-b)_0$. An estimate of $(b-y)_0$ is found by substituting $c_1$ for $c_0$ in the formula for $(b-y)_0$ in section three of \cite{shob84}. We estimate the reddening $E(b-y)=(b-y)-(b-y)_0$ and the quantity $c_0=c_1-0.19\:E(b-y)$. We then compute an improved value of $(b-y)_0$, and repeat the process until the value of $(b-y)_0$ changes by less than 0.001 mag from the previous iteration. With this process, we find $E(b-y)=0.22\pm0.03$ for TT Her. Since $E(b-y)=0.74\:E(B-V)$ \citep{craw75}, $E(B-V)=0.30\pm0.03$ and using Landolt's observed $B-V$, we find $(B-V)_0=0.02\pm0.03$, a value consistent with an A0 to A2 classification.

\subsection{Light and Radial Velocity Curve Analysis}

We analyzed our radial velocity and photometric observations simultaneously using the 2013 version of the Wilson-Devinney (WD) program \citep{wd71,rew79,rew90,rew12}. The weights for each light and radial velocity curve were determined by allowing WD to adjust the weights based on the scatter in the data. Initial estimates of the weights are made by measuring the scatter in representative sections of the light or velocity curves, such as the maxima of a light curve, and then for each subsequent iteration WD will compute standard deviations for each data curve and apply the appropriate weight automatically. Because of the lack of agreement on the spectral type of the primary by previous observers, we decided to explore solutions to our data for both early and late-type primaries. Previous analyses of TT Her resulted in a variety of morphologies: detached \citep{kal85}, semi-detached \citep{kal85,mil89}, and overcontact \citep{kwee83}. 

\subsubsection{Solutions Assuming a Late-type Primary}

We began our fitting experiments assuming a late-type primary since a number of observers had given the system late-A or early-F classifications. The first solutions were done with a mean effective temperature for the primary of $T_1=7300$ K, corresponding to a spectral type of F0. The initial manual fit to the light and radial velocity curves was done with a detached configuration (WD mode 2) and then the differential corrections (DC) program of WD was used to adjust the fit. Certain parameters, such as gravity darkening exponents and bolometric albedoes, were fixed at their theoretical values. The temperatures of the stars were always such that values appropriate for radiative envelopes were assumed for the primary and convective envelopes for the secondary. WD can treat limb darkening with a variety of limb darkening laws with either fixed coefficients, or it can locally interpolate in effective temperature and (log) surface gravity for the coefficients. We used the latter approach with a square root limb darkening law. We also adjusted third light in our solutions but never found any values statistically distinguishable from zero. 

The solution quickly moved toward a semi-detached configuration with the secondary star filling its Roche lobe and the primary close to, but not quite filling its Roche Lobe. Configurations such as this are known to require a detailed treatment of the reflection effect \citep{rew90} for the highest accuracy, and we used the detailed reflection option in WD for all of the fitting experiments described herein. 

We also attempted to find an overcontact solution (WD mode 3) where both surface potentials are constrained to be equal, but found that the solution would always move towards a state where the system was not in contact. In that configuration, there would be no physical reason for the surface potentials to be equal, and thus we concluded that an overcontact configuration was unrealistic. The large temperature difference between the two stars also argues against an overcontact configuration.

\begin{figure}
\epsscale{1.0}
\plotone{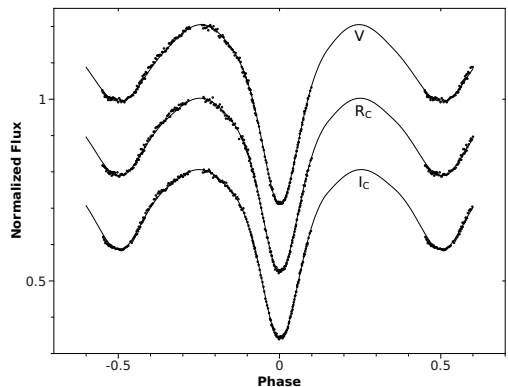}
\caption{The fit to the light curves for a late-type primary with $T_1=7300$ K and the system in a semi-detached configuration with the secondary star filling its Roche lobe. Stellar atmosphere models were used for both stars.
\label{latefit}}
\end{figure}

\begin{figure}
\epsscale{1.0}
\plotone{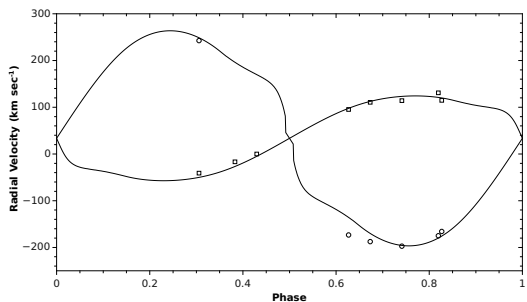}
\caption{The fit to the radial velocity curves for a late-type primary with $T_1=7300$ K and the system in a semi-detached configuration with the secondary star filling its Roche lobe. The observed radial velocities of the primary are shown as squares and those of the secondary as circles. The standard errors of the radial velocities are about the same size as the data markers.
\label{latefit_rv}}
\end{figure}

Our light curves of TT Her show a mild asymmetry between the two maxima, with the maximum preceeding the primary minimum slightly higher. Rather than trying to model spots, we took the approach of eliminating the data for one maximum and then finding the solution with DC. Then we repeated the solution with the other maximum in place. We found that using the data in the maximum preceding primary eclipse worked best, in that we got good fits to it in all three filters. The maximum following primary eclipse showed fitting problems, mainly with the $V$ curve. This maximum also shows enhanced scatter compared to the other one. For example, in the $V$ curve, the scatter in the maximum preceding primary eclipse is 0.006 mag while the scatter in the other is 0.009 mag. This increased scatter in the one maximum is also found in the \cite{kwee83} and the \cite{burchi82} light curves. The latter observations also seem to show a wavelength dependence of this disparity in the scatter of the two maxima, with a greater difference in the scatter at shorter wavelengths. We therefore concluded that the maximum following primary eclipse was the one being affected, perhaps by spot activity on the secondary star or absorption by matter streams, and subsequent DC solutions were done with the data for phases 0.1 to 0.45 excluded. Figure \ref{latefit} shows the fits to the light curves to be quite satisfactory, with the scatter of the residuals being 0.006, 0.006 and 0.005 mag in the $V$, $R_C$ and $I_C$ passbands, thus similar to the observational scatter of the data. (We note that the average residual is about ten times larger when we include the photometric data between phases 0.1 and 0.45.) However, the fits to the radial velocity curves shown in Figure \ref{latefit_rv} do not look very good, especially for the secondary star. The average residual of the fit to the primary star is 7.9 km\: sec$^{-1}$ and 19.6 km\: sec$^{-1}$ for the secondary star. The poor fit to the radial velocity data is an important clue to uncovering the true nature of TT Her.

It has long been known that photometry can yield accurate mass ratios in certain situations. \cite{terrell05} show how semi-detached and overcontact systems with complete eclipses can yield accurate photometric mass ratios. Our semi-detached solution does have complete eclipses, and thus we expect that the DC program will strongly prefer a certain photometric mass ratio ($q_{ptm}$), which should, assuming the velocities are accurate, match the spectroscopic mass ratio ($q_{sp}$). In a semi-detached configuration, the spectra might be affected by mass transfer activity. Our spectra, although quite limited in wavelength coverage, do not show any noticeable emission features, so we have no reason to suspect that the velocities are affected by circumstellar emission and accept them as accurate measurements of the motions of the two stars. Figure \ref{latefit_rv} and the large average residuals clearly show that the radial velocities are not being predicted as accurately as they should be. Visual inspection of Figure \ref{latefit_rv} also shows that the mass ratio might be lower than the semi-detached model predicts since a lower mass ratio would decrease the amplitude of the primary's radial velocity curve and increase that of the secondary. We did a solution to just the radial velocities assuming the stars were point masses, and the resulting mass ratio was about $0.37\pm0.02$. The fact that the photometric data are pushing the solution towards a different mass ratio is an indication that the model is deficient in some way, again assuming that the velocities are accurate. Given that our velocities match those of \cite{san37} reasonably well, we were inclined to believe that there was an issue with the semi-detached model. We soon discovered that this was indeed the case.

The biggest problem with this solution, however, is that it does not make sense astrophysically. The masses for the two components from this solution are about 2.7 $M_\sun$ and 1.1 $M_\sun$, values much too high for these effective temperatures. We then decided to look at the possibility that the primary star was indeed of an earlier spectral type.

\subsubsection{Solutions Assuming an Early-type Primary}
\label{sec:solution}

To test models for an early-A primary, we set $T_1=9500$ K and again started from a detached configuration. As before, the solution moved towards a semi-detached configuration with the secondary filling its Roche lobe. However, in this case the fit to the light curves was not very good at the bottoms of the eclipses, especially the secondary eclipse but also in the primary eclipse in the $I_C$ filter. Suspecting that the problem might be that the solution was stuck in a local minimum, we tried a wide variety of starting points in parameter space, but the same minimum was always recovered.

\begin{figure}
\epsscale{1.0}
\plotone{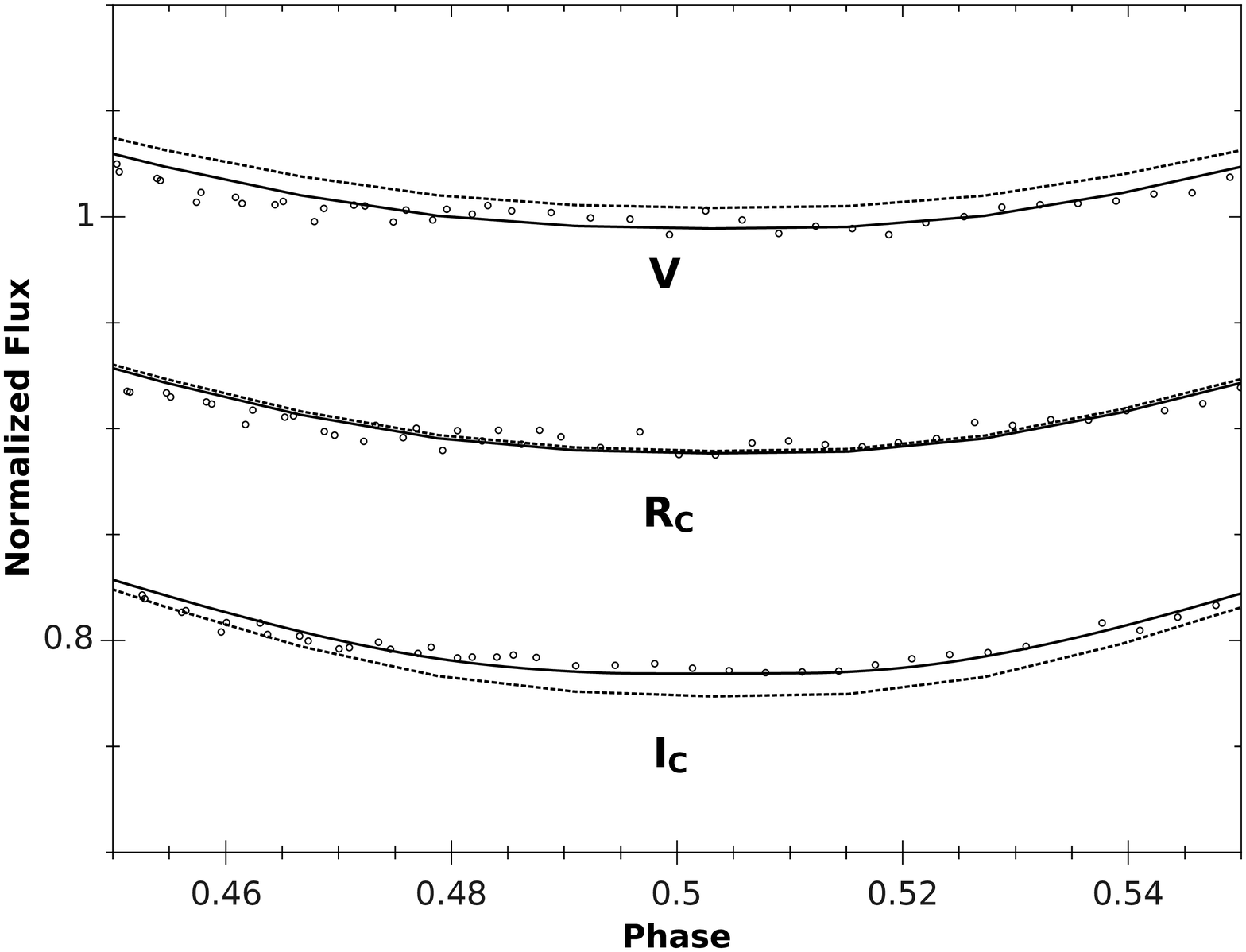}
\caption{The fits to the bottoms of the secondary eclipses for a semi-detached configuration and $T_1=9500$ K when using a blackbody model for the primary star's radiation (solid curves) and a stellar atmosphere model (dashed curve).\label{sdfit}}
\end{figure}

Various attempts to address these fitting issues, such as a different limb darkening law or different metallicities, showed no appreciable improvement. One change did, however, result in a noticeably better fit and that was to use a blackbody rather than stellar atmospheres to model the radiation of the primary star. Figure \ref{sdfit} shows the improved fit of this approach (solid curves) over that of using stellar atmospheres (dashed curves). The fits to both eclipses in all three filters are very good.

While the fits to the photometry looked reasonably good, the fits to the radial velocity curves showed the same problems as the semi-detached solution for a late-type primary. The computed radial velocity curves looked indistinguishable from those in Figure \ref{latefit_rv} and the average residuals were very similar. The photometry seemed to be driving the solution to a slightly higher mass ratio (about 0.40) than the radial velocities supported (about 0.37). The model still seemed deficient in some way.

As \cite{terrell05} discuss for semi-detached systems, the lobe-filling constraint on one star, coupled with the accurate relative radius determination from complete eclipses leads to an accurate determination of the photometric mass ratio, $q_{ptm}$. The DC algorithm zeroes in on the same value of $q_{ptm}$ for TT Her from a variety of starting points in parameter space, so we see the expected strongly convergent behavior. The question is why it converges to a value different from $q_{sp}$. The answer must lie in the fact that the relative radius of the secondary is converging to a value that leads to the inconsistency in the photometric and spectroscopic mass ratios. In order to have a lower $q_{ptm}$, the secondary star needs to be smaller (see figure 2 in \cite{terrell05}), thus requiring a larger primary in order to keep the sum of the relative radii the same. Since the secondary is constrained to fill the lobe, the only way to address the problem is in adjusting the relative radius of the primary. The depths of the eclipses are, however, fit very nicely, so simply changing the surface potential of the primary cannot solve the problem because then the eclipse depths change. Clearly what is needed is a change in the shape of the primary star that doesn't alter the projected area during eclipses but makes the star wider, and that can be accomplished with faster than synchronous rotation.

\begin{figure}
\epsscale{1.0}
\plotone{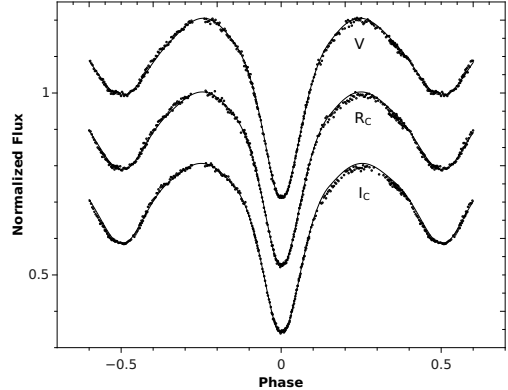}
\caption{The fit to the light curves for an early-type primary with $T_1=9500$ K and the system in a double contact configuration. All photometric observations are shown to illustrate the light curve asymmetries, but the observations between phases 0.1 and 0.45 were excluded from the fit, as shown in Figure \ref{latefit}. The primary star was modeled as a blackbody while the secondary was modeled with stellar atmopsheres.
\label{earlyfit}}
\end{figure}

\begin{figure}
\epsscale{1.0}
\plotone{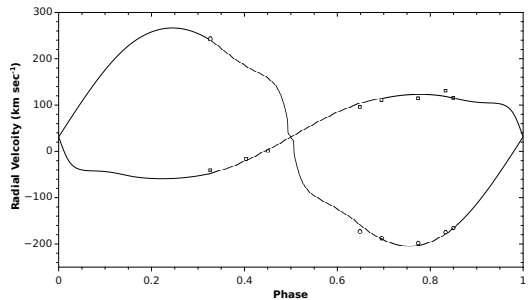}
\caption{The fit to the radial velocity curves for an early-type primary with $T_1=9500$ K and the system in a double contact configuration. The observed radial velocities of the primary are shown as squares and those of the secondary as circles. The standard errors of the radial velocities are about the same size as the data markers. Note the significantly improved fit to the radial velocities of both stars as compared to the semi-detached solution in Figure \ref{latefit_rv}. \label{earlyfit_rv}}
\end{figure}

We then allowed DC to adjust the rotation parameter ($F_1$) for the primary star and it quickly reached a configuration where the primary was filling its critical surface, making the system a double contact binary as defined by \cite{rew79}. DC converged rather quickly on a solution, with the primary rotating at $1.25\pm0.01$ times the synchronous value, that fit the light (Figure \ref{earlyfit}) and the radial velocity curves (Figure \ref{earlyfit_rv}) quite nicely. The average residual of the fit to the primary star's radial velocity is 7.2 km\: sec$^{-1}$ and 7.5 km\: sec$^{-1}$ for the secondary star's radial velocities, both being improvements from the values for the solution assuming a late-type primary. The mass ratio converged to a value of $0.383\pm0.002$, which is in good agreement with the spectroscopic mass ratio we found earlier, $0.37\pm0.02$. The values of the adjusted parameters from the solution are given in Table 3.

Since our observations spanned about three years and TT Her is known to show period variations, we tried adjusting the orbital period derivative in addition to the orbital period and the reference epoch, but were unable to determine a reliable value. \cite{kre08} show that the eclipse timing diagram can be reasonably well fit by a secular period decrease plus an approximately 40-year periodic term due to the influence of a third star. This secular period decrease could imply mass transfer from the rapidly rotating primary star back to the secondary star. If that is happening, the matter stream would be projected against the disk of the secondary near orbital phase 0.25. Since the light curve maximum at that phase is the one that shows enhanced scatter and fitting problems, it is consistent with this mass transfer scenario. A semi-detached model with the secondary filling the Roche lobe would not present an obvious explanation for a matter stream flowing from the primary to the secondary. We did perform a solution like that of \cite{mil89} where the primary was filling its Roche lobe (and rotating synchronously) and the secondary's surface potential was allowed to vary, i.e. a reverse Algol \citep{leung89}, but the fit to the radial velocities was visibly inferior to the double contact solution, with average residuals for the primary and secondary radial velocities of 17.2 km\: sec$^{-1}$ and 11.0 km\: sec$^{-1}$ respectively. The mass ratio in this case was even higher than in the semi-detached solution, about 0.46.

\begin{deluxetable}{cc}
\tablecaption{Final Solution Parameters for the Double Contact Model of TT Her}
\tablenum{3}
\tablehead{
   \colhead{Parameter} & \colhead{Value}
} 
\startdata
$a\: (R_\odot)$ & 6.23 $\pm$ 0.10 \\
$F_1$ & 1.25 $\pm$ 0.01 \\
$V_\gamma\: (km\: sec^{-1})$ & 31 $\pm$ 2 \\
$i\: (^\circ)$ & 82.72 $\pm$ 0.04 \\
$T_2\: (K)$ & 5750 $\pm$ 9 \\
$q\: (M_2/M_1)$ & 0.383 $\pm$ 0.002 \\
$HJD_0$ & 2454603.7445 $\pm$ 0.0001 \\
$P\: (days)$ & 0.912099 $\pm$ 0.000005 \\
$\frac{L_1}{L_1+L_2}\: (V)$ & 0.931 $\pm$ 0.001 \\
$\frac{L_1}{L_1+L_2}\: (R_C)$ & 0.916 $\pm$ 0.001 \\
$\frac{L_1}{L_1+L_2}\: (I_C)$ & 0.900 $\pm$ 0.001 \\
\enddata
\label{paramtable}
\end{deluxetable}

In addition to the better fits to the observations, the double contact solution with an early-type primary has the advantage that the measured masses for the stars are very close to what would be expected for the two stars. The primary's mass is $2.82 \pm 0.14 M_\sun$, consistent with an A0 spectral type and the assumed $T_1=9500\: K$. The secondary's mass is $1.08 \pm 0.05 M_\sun$. Double contact binaries are expected to arise from mass transfer, where the envelope of the primary has been spun up after the system passed through the rapid phase of mass transfer \citep{rew85,rew89}. If that is the case in TT Her, we would expect the secondary to be the more evolved star, and that is indeed the case. The mean radius of the secondary is $1.86 \pm 0.03 M_\sun$, yielding a $log\: g$ value of $3.93 \pm 0.06$, thus showing significant evolution for a solar-type star. With a derived mean radius of $2.79 \pm 0.04$, the $log\:g$ value for the primary is $4.00 \pm 0.06$, making it only slightly evolved. 

\section{Conclusions}

New light and radial velocity curves of TT Her have been analyzed and indicate that the system is in a double contact configuration. The double contact model fits the observations best, and gives a consistent evolutionary view of the system. The only troublesome issue with this solution is the poorer fit to the light curves for a stellar atmosphere model than a blackbody for the primary star, and may indicate something unusual about the star that more detailed spectra than ours might reveal. Given the widely varying spectral types assigned to the star by previous researchers, our finding may simply be another manifestation of that unusual nature, rather than indicating a fundamental problem with our solution. This, along with the rare double contact morphology, is a strong motivation for further observations of the system.

A double contact configuration indicates that the system has undergone large-scale mass transfer, and the envelope of the primary has been spun up by mass transfer from the secondary. The surface gravities of the two stars are consistent with this evolutionary scenario. Whereas semi-detached solutions (with either the primary or the secondary star filling the Roche lobe) always show a discrepancy between $q_{ptm}$ and $q_{sp}$, the double contact solution does not, leading us to conclude that a double contact model finally provides a good solution for the morphology of this previously vexing system. With such a short orbital period, tidal forces will rapidly synchronize the rotation of the primary, putting TT Her in a short-lived, and thus very rare, configuration. Detailed evolutionary models, constrained by our results on the properties of the stars, might prove very rewarding. 

\acknowledgments
It is a pleasure to thank the staff members at the DAO (especially Dmitry Monin and Les Saddlemyer) for their usual splendid help and assistance. This research has made use of the SIMBAD database, operated at CDS, Strasbourg, France. We also thank Prof. R.E. Wilson for comments on an early version of the manuscript, and the referee for a very helpful review.

\end{document}